# Longitudinal dispersion of DNA in nanochannels

Theo Odijk

*Complex Fluids Theory, Faculty of Applied Sciences, Delft University of Technology, Julianalaan 67, 2628 BC Delft, The Netherlands*

E-mail: odijktcf@wanadoo.nl

**Abstract**

A theory is presented of the longitudinal dispersion of DNA under equilibrium confined in a nanochannel. Orientational fluctuations of the DNA chain build up to give rise to substantial fluctuations of the coil in the longitudinal direction of the channel. The translational and orientational degrees of freedom of the polymer are described by the Green function satisfying the usual Fokker-Planck equation. It is argued that this is analogous to the transport equation occurring in the theory of convective diffusion of particles in pipe flow. Moreover, Taylor's method may be used to reduce the Fokker-Planck equation to a diffusion equation for long DNA although subtleties arise connected with the orientational distribution of segments within the channel. The longitudinal "step length" turns out to be proportional to the typical angle of a DNA segment to the sixth power. The dispersion is underestimated compared to experiment, probably because the harmonic approximation is used to describe the polymer confinement.

## Introduction

The mass production of nanostructures has become routine. Since nanoconfined DNA is of obvious interest in fields ranging from analytical chemistry to genomics, there has been considerable activity in DNA nanoscience in the past decade. In particular, research on the behavior of DNA enclosed in nanochannels and nanoslits has been expanding rapidly (see e.g. recent reviews [1-6]).

The segment distribution of nanoconfined DNA is strongly inhomogeneous which leads to taxing statistical mechanical problems. Orientational and translational degrees of freedom are strongly coupled in the Fokker-Planck equation describing the probability [7]. Although this type of coupling has been known for some time in various transport problems including chemical rate theory [8], convective diffusion [9-11] and the diffusion of neutrons in nuclear reactors [12,13], little progress in the practical solution of the Fokker-Planck equation seems to have been achieved since then.

The myriad orientational and transverse fluctuations of a wormlike chain in a nanotube ultimately add up to significant fluctuations of the worm in the longitudinal direction. If the tube is thin enough, the chain of contour length L may be viewed as a sequence of $L/\lambda$ deflection segments of length $\lambda = P\tilde{\theta}^2$ where $P$ is the persistence length and $\tilde{\theta}$ is the typical angle of a segment with respect to the channel axis [14]. The fluctuation of such a segment in the longitudinal direction is of order $\lambda\tilde{\theta}^2$. There are $L/\lambda$ of such random displacements so that the longitudinal dispersion, that is the total mean-square displacement $\langle\Delta z^2\rangle$ is of order $(L/\lambda)\lambda^2\theta^4$. Thus a simple scaling analysis establishes a remarkable sixth power law in terms of the angle $\tilde{\theta}$. But of considerable relevance is the precise magnitude of this effect whose computation will be attempted here. The dispersion is important in gauging the accuracy of locating genes on the DNA by fluorescent microscopy [15,16].

The orientation of the wormlike DNA couples to the spatial gradient in the Fokker-Planck equation for the probability. If the contour of the DNA is viewed as "time", this term in the equation may be viewed as "convective". It is then tempting to borrow Taylor's trick from the theory of convective diffusion [10] to eliminate the orientational degrees of freedom so as to reduce the Fokker-Planck equation to the "diffusion" equation in the longitudinal direction. Nevertheless, the analogy is not straightforward because the orientation of the chain is not an independent variable. It turns out to be useful to check the analysis by evaluating the second moment $\langle\Delta z^2\rangle$ via an entirely independent route. It is difficult to account rigorously for the boundary condition on the probability owing to the hard channel walls. I adopt a harmonic approximation as discussed by Burkhardt [17] which, if suitably rescaled, appears to provide an accurate representation of the actual physical problem [15,18].

## Fokker–Planck Equation

The DNA coil is represented by a wormlike chain of contour length $L$ and persistence length $P$. It is confined in a nanochannel of square cross-section whose sides are $A$ long. I focus on the deflection regime $(A<P)$ under circumstances where backfolding or hairpin formation is absent [19]. Cartesian coordinates are introduced where the centerline of the nanochannel is the $z$ axis and the $x$ and $y$ axes are parallel to the sides of its cross-section. The configuration of the chain is determined by the radius vector $\vec{r}(s) = (x(s), y(s), z(s))$ as a function of the contour distance $s$ from one end. But the statistical properties of the polymer also depend on the unit vector $\vec{u}(s) = d\vec{r}/ds$ as exemplified by the Green function $G(\vec{r},\vec{u};\vec{r}_0,\vec{u}_0;L)$ which satisfies a well-known Fokker-Planck equation [7]

$$\frac{\partial G}{\partial L} + \vec{u}\cdot\frac{\partial G}{\partial \vec{r}} - \frac{1}{2P}\Delta_{\vec{u}}G = 0 \qquad (1)$$

$$\lim_{L\to 0} G = \delta(\vec{r}-\vec{r}_0)\delta(\vec{u}-\vec{u}_0) \qquad (2)$$

where $\Delta_{\vec{u}}$ is the Laplacian on the unit sphere. The chain starts at $\vec{r}_0$ with orientation $\vec{u}_0$ and ends at $\vec{r}$ with orientation $\vec{u}$; it is well to recall that $G$ is a conditional probability.

The DNA chain cannot cross the walls of the nanochannel nor can it bend discontinuously. If $\vec{n}$ is the inward pointing unit vector perpendicular to the channel walls, the boundary conditions on the Green function must be that $G$ vanishes for $\vec{u}\cdot\vec{n}>0$ as $\vec{r}$ approaches a wall but equal to some constant to be determined in case $\vec{u}\cdot\vec{n}<0$. Nevertheless, in the main, the chain is confined close to the centeraxis of the channel so it has proven useful to replace the mathematically difficult problem of a worm confined by hard walls by the simpler one of a chain confined harmonically in a potential $\frac{1}{2}b(x^2+y^2)$ along the $z$ axis. Burkhardt solved eq (1) with this potential added [17], though he discarded the $z$ dependence which he did not need in his analysis to the leading order. By a suitable matching of the two respective problems [15], an analytical estimate for the elongation of the DNA may be derived which is very close to a recent numerical result [18]. Thus, the original problem may be replaced by an appropriate harmonic approximation, a caveat I will use to justify several steps in what follows.

In the deflection regime, the angle $\vec{\theta}(s)=(\theta_x(s),\theta_y(s))$ between the chain tangent and the $z$ axis is small so that the tangential unit vector is given by $\vec{u}=(\theta_x,\theta_y,1-\frac{1}{2}\theta^2)$. Hence, eq(1) simplifies to

$$\frac{\partial G}{\partial L}+\left(1-\tfrac{1}{2}\theta^2\right)\frac{\partial G}{\partial z}+\theta_x\frac{\partial G}{\partial x}+\theta_y\frac{\partial G}{\partial y}-\frac{1}{2P}\frac{\partial^2 G}{\partial\theta_x^2}-\frac{1}{2P}\frac{\partial^2 G}{\partial\theta_y^2}=0 \quad (3)$$

It is now convenient to interpret this expression as describing "convective diffusion" where $L$ is "time", $\tfrac{1}{2}P$ is a "diffusion coefficient" and $\vec{u}$ is a "velocity" of "particles". At this stage, I introduce an angular distribution which is uniform between the extrema $-\vec{\theta}_m$ and $\vec{\theta}_m\equiv(\theta_m,\theta_m)$ which will be discussed below. The average "velocity" is $1-\tfrac{1}{2}\overline{\theta^2}$ and it is expedient to view the diffusional process within this frame. Furthermore, let us define

$$H(z,\theta_x,\theta_y;z_0;L)\equiv \iiiint\int dxdydx_0dy_0d\vec{u}_0\, G \quad (4)$$

across the channel so that eq(3) reduces to

$$\frac{\partial H}{\partial L}+\omega\frac{\partial H}{\partial z}-\frac{1}{2P}\frac{\partial^2 H}{\partial\theta_x^2}-\frac{1}{2P}\frac{\partial^2 H}{\partial\theta_y^2}=0 \quad (5)$$

I have introduced a rescaled velocity $\omega\equiv\tfrac{1}{2}\overline{\theta^2}-\tfrac{1}{2}\theta^2$. In the integration of eq(3) over $x$ and $y$, the terms at the channel walls are regarded as negligible in accordance with the harmonic approximation (in the latter the integrations are over all space).

**Diffusion Equation**

The reduced Fokker-Planck eq (5) is quite similar to that occurring in the theory of convective diffusion of particles introduced into the flow within a cylindrical pipe. The shearing flow greatly enhances the bare diffusion because the particles do not move at the wall of the pipe owing to the stick boundary condition pertaining to the fluid whereas the particles are swept away fast by the fluid at the centerline [10]. Nevertheless, the analogy is not exact since the flow is independent of the particle concentration in the latter case; in the polymer problem the distribution of the "velocity" $\omega$ depends implicitly on the probability function $H$ which needs to be determined (see below).

Taylor solved the particle diffusion problem by an iterative Ansatz [10,11] which turns out to be convenient also for the DNA problem at hand. At long "times" $L$, the $\theta$ dependence in eq (5) may be viewed locally so that $\partial H/\partial z$ is approximately constant and $\partial H/\partial L\simeq 0$. Under the same initial conditions the initial behavior of $H$ may be considered smoothed out implying that $H$ could be close to the orientational average $\overline{H}$. Thus we first solve

$$2P\omega\frac{\partial\overline{H}}{\partial z}-\frac{\partial^2 H}{\partial\theta_x^2}-\frac{\partial^2 H}{\partial\theta_y^2}=0 \quad (6)$$

which yields

$$H=P\frac{\partial\overline{H}}{\partial z}\left(B+\tfrac{1}{2}\theta^2\overline{\theta^2}-\tfrac{1}{12}\theta^4\right) \quad (7)$$

where $B$ is a constant and a term linear in $\vec{\theta}$ is absent in view of symmetry. The constant is eliminated by averaging eq (7)

$$H = \overline{H} + P\frac{\partial \overline{H}}{\partial z}\left(\tfrac{1}{2}\theta^2\overline{\theta^2} - \tfrac{1}{2}\left(\overline{\theta^2}\right)^2 \right.$$
$$\left. -\tfrac{1}{12}\theta^4 + \tfrac{1}{12}\overline{\theta^4}\right) \quad (8)$$

Note that, at large $z$, $H$ is indeed close to $\overline{H}$ so that the approximation scheme is self-consistent. In the diffusion analogy the "flux" here is the "velocity" $\omega$ times the "concentration" $H$ integrated over the "cross section" divided by the area. The effective "diffusion coefficient" $D_{eff}$ equals minus the "flux" divided by the gradient $\partial \overline{H}/\partial z$

$$D_{eff} = \tfrac{1}{4}P\overline{\left(\theta^2 - \overline{\theta^2}\right)\left(\theta^2\overline{\theta^2} - \left(\overline{\theta^2}\right)^2 - \tfrac{1}{6}\theta^4 + \tfrac{1}{6}\overline{\theta^4}\right)}$$
$$= \tfrac{1}{4}P\left(\tfrac{7}{6}\overline{\theta^2}\ \overline{\theta^4} - \left(\overline{\theta^2}\right)^3 - \tfrac{1}{6}\overline{\theta^6}\right) \quad (9)$$
$$= \tfrac{1}{18}P\left(\overline{\theta^2}\right)^3$$

Here, the last simplification arises because the orientation is uniform. We therefore end up with a "diffusion" equation

$$\frac{\partial \overline{H}}{\partial L} = D_{eff}\frac{\partial^2 \overline{H}}{\partial z^2} \quad (10)$$

valid at large $L$ and $z$, which may also be derived by averaging eq (5).

Eq(10) is only intermediate since $\overline{H}$ has to be connected to the Green function $\overline{G} = \int d\vec{\theta}\ H$ integrated over the entire angular half-space and we also need the second moment $\overline{\theta^2}$. These two issues are now settled within the harmonic approximation. For a long chain, Burkhardt has shown that the original Green function neglecting the $z$ dependence satisfies [17]

$$G_h \sim e^{-\mu_0 L}\ \Psi_0(x, y, \theta_x, \theta_y)\Psi_0(x, y, -\theta_x, -\theta_y) \quad (11)$$

$$\Psi_0 \sim \exp\left[-\tfrac{1}{\sqrt{2}}\left(\hat{x}^2 + \hat{y}^2 + \hat{\theta}_x^2 + \hat{\theta}_y^2\right)\right.$$
$$\left. + \hat{x}\hat{\theta}_x + \hat{y}\hat{\theta}_y\right] \quad (12)$$

in terms of appropriately scaled variables $(\hat{x}, \hat{y}) = P^{1/8}b^{3/8}(x, y)$ and $(\hat{\theta}_x, \hat{\theta}_y) = P^{3/8}b^{1/8}(\theta_x, \theta_y)$. Hence, the orientational distribution $\int dx \int dy\ \Psi_+\Psi_-$ is precisely a Gaussian. Accordingly, the uniform distribution chosen earlier to simulate the actual angular distribution of segments within the nanochannel is now taken to be Gaussian in the harmonic approximation. It is then plausible to set the respective second moments equal so that $\theta_m^2 = 2\overline{\theta^2} = 2^{1/2}b^{-1/4}P^{-3/4}$ in terms of the harmonically computed moment which will be discussed below. In a similar vein, one can argue that $\overline{H}$ is proportional to $\overline{G}$ in view of the scaling structure of the harmonic approximation. Thus, I conclude that a coarse-grained version of eq (1) satisfies

$$\frac{\partial \overline{G}}{\partial L} = \tfrac{1}{18}P\left(\overline{\theta^2}\right)^3\frac{\partial^2 \overline{G}}{\partial z^2} \quad (13)$$

From the analysis above, it is not clear how good is the estimate we have for the small numerical coefficient appearing in eq (13). A second independent calculation of this is given in the next section.

**Mean-Square Longitudinal Dispersion**

Although eq (13) has been divided at large $z$, we know that the Green function $\overline{G}$ must satisfy the initial condition

$$\lim_{L\to 0}\overline{G} = \delta(z - z_0) \quad (14)$$

Thus, $\overline{G}$ at large $z$ is simply a Gaussian

$$\overline{G} = \frac{1}{2(\pi D_{eff}L)^{1/2}}\exp\left[-\frac{(z - \langle z\rangle)^2}{4D_{eff}L}\right] \quad (15)$$

with mean-square dispersion

$$\langle \Delta z^2\rangle \equiv \langle(z - \langle z\rangle)^2\rangle = \langle z^2\rangle - \langle z\rangle^2 = 2D_{eff}L \quad (16)$$

Here, $\langle \ \rangle$ denotes an average over all configurations of the chain and we have reverted to the original reference frame. The left-hand side of eq (16) may be rewritten in terms of an orientational correlation function since one has

$$z(L) = \int_0^L ds \ \cos\theta(s)$$
$$= L - \tfrac{1}{2}\int_0^L ds \ \theta^2(s) + \tfrac{1}{6}\int_0^L ds \ \theta^4(s) - \ldots\ldots \quad (17)$$

To the leading order and at large L, this yields

$$\langle \Delta z^2 \rangle = \tfrac{1}{4}\int_0^L ds \int_0^L dt \left( \langle \theta^2(s)\theta^2(t) \rangle - \langle \theta^2(s) \rangle \langle \theta^2(t) \rangle \right) \quad (18)$$
$$= \tfrac{1}{4} L \overline{(\theta^2)}^2 \int_0^\infty dp \ (Q(p)-1)$$

where the correlation function

$$Q(p) \equiv \frac{\langle \theta^2(0)\theta^2(p) \rangle}{\langle \theta^2 \rangle^2} \quad (19)$$
$$= \frac{\langle \theta^2(0)\theta^2(p) \rangle}{\left(\overline{\theta^2}\right)^2}$$

Accordingly, we need to compute $Q(p)$ in the harmonic approximation.

In order to evaluate the average of some quantity $h(\vec{\theta}_s, \vec{\theta}_t)$ one needs the complete Green function including the translational degrees of freedom because the differential operator in eq (1) is non-selfadjoint

$$\langle h(\vec{\theta}_s, \vec{\theta}_t) \rangle =$$
$$\frac{\iint d\vec{r}_0 d\vec{\theta}_0 \iint d\vec{r}_s d\vec{\theta}_s \iint d\vec{r}_t d\vec{\theta}_t \iint d\vec{r}_n d\vec{\theta}_n \ h(\vec{\theta}_s, \vec{\theta}_t) \ G(\vec{r}_0, \vec{\theta}_0; \vec{r}_s, \vec{\theta}_s; s) \ G(\vec{r}_s, \vec{\theta}_s; \vec{r}_t, \vec{\theta}_t; t-s) \ G(\vec{r}_t, \vec{\theta}_t; \vec{r}_n, \vec{\theta}_n; L-t)}{\iint d\vec{r}_0 d\vec{\theta}_0 \iint d\vec{r}_n d\vec{\theta}_n \ G(\vec{r}_0, \vec{\theta}_0; \vec{r}_n, \vec{\theta}_n; L)}$$
$$= \iint d\vec{r}_s d\vec{\theta}_s \iint d\vec{r}_t d\vec{\theta}_t \ h(\vec{\theta}_s, \vec{\theta}_t) e^{\mu_0(t-s)} \Psi_0(\vec{r}_s, -\vec{\theta}_s) G(\vec{r}_s, \vec{\theta}_s; \vec{r}_t, \vec{\theta}_t; t-s) \Psi_0(\vec{r}_t, \vec{\theta}_t)$$

(20)

The second equality is valid in the limit of very long chains where ground state dominance applies: $G(\vec{r}_0, \vec{\theta}_0; \vec{r}_n, \vec{\theta}_n; L) \sim (\exp -\mu_o L) \Psi_0(\vec{r}_0, \vec{\theta}_0) \Psi_0(\vec{r}_n, -\vec{\theta}_n)$.

Omitting the $z$ dependence in eq (20) which merely gives rise to higher order terms, I next use Burkhardt's expression [17] for the Green function in the harmonic approximation in terms of the scaled variables introduced earlier.

$$G_h(\vec{r}, \vec{\theta}; \vec{r}_0, \vec{\theta}_0; L) = NPb(s_h^2 - s^2)^{-1} \exp -S(\vec{r}, \vec{\theta}; \vec{r}_0, \vec{\theta}_0; L) \quad (21)$$

$$S(\vec{r}, \vec{\theta}; \vec{r}_0, \vec{\theta}_0; L) \equiv \tfrac{1}{2}(s_h^2 - s^2)^{-1}$$
$$\left\{ \sqrt{2}(c_h s_h + cs)(\hat{\vec{r}}^2 + \hat{\vec{r}}_0^2) + \sqrt{2}(c_h s_h - cs)(\hat{\vec{\theta}}^2 + \hat{\vec{\theta}}_0^2) \right.$$
$$-2(s_h^2 + s^2)(\hat{\vec{r}} \cdot \hat{\vec{\theta}} - \hat{\vec{r}}_0 \cdot \hat{\vec{\theta}}_0) - 2\sqrt{2}(c_h s + s_h c)\hat{\vec{r}} \cdot \hat{\vec{r}}_0$$
$$\left. +2\sqrt{2}(c_h s - s_h c)\hat{\vec{\theta}} \cdot \hat{\vec{\theta}}_0 - 4s_h s (\hat{\vec{r}} \cdot \hat{\vec{\theta}}_0 - \hat{\vec{\theta}} \cdot \hat{\vec{r}}_0) \right\}$$

(22)

where $c_h \equiv \cosh(\hat{L}/\sqrt{2})$, $S_h \equiv \sinh(\hat{L}/\sqrt{2})$, $c \equiv \cos(\hat{L}/\sqrt{2})$, $s \equiv \sin(\hat{L}/\sqrt{2})$, $\hat{L} \equiv b^{1/4} P^{-1/4} L$ and $N$ is a dimensionless normalization constant. Note that eq (21) and eq (22) reduce to eq (11) and eq (12) at large $L$. The correlation function $Q(p)$ is computed with the help of eq (19-22) and depends only on the scaled variable $\hat{p} = b^{1/4} P^{-1/4} p = p/(\sqrt{2} \ \overline{\theta^2} P)$ where $\overline{\theta^2} P$ is the deflection length so that

$$\langle \Delta z^2 \rangle = \frac{\sqrt{2}}{4} I (\overline{\theta^2})^3 PL \quad (23)$$

$$I = \int_0^\infty d\hat{p}\,(Q(\hat{p}) - 1) \qquad (24)$$

Hence the mean-square dispersion indeed scales exactly as that in the analysis of the Fokker-Planck equation (see eq (16)). The integrations over the translational and orientational variables in eq(20) are straightforward but lead to extremely tedious expressions which are not reproduced here. Nevertheless, as pointed out to me by Peter Prinsen, a useful series expansion of $Q(\hat{p})$ is readily derived from these with the help of Mathematica

$$Q(\hat{p}) = 2 - 2\sqrt{2}\hat{p} + 3\hat{p}^2 - \sqrt{2}\hat{p}^3 + \frac{\hat{p}^4}{6} + \frac{\hat{p}^5}{5\sqrt{2}} - \frac{7\hat{p}^6}{90} + \frac{\hat{p}^7}{45\sqrt{2}} + \frac{\hat{p}^8}{2520} - \frac{17\hat{p}^9}{11340\sqrt{2}} + O(\hat{p}^{10}) \qquad (25)$$

This allows us to check the numerical integration of $I = 0.2875$ for the full expression for $Q(\hat{p})$ turns out to be computationally suspect as $\hat{p}$ approaches zero.

*Discussion*

The "effective diffusion coefficient" computed via the second moment by brute force is given by

$$D_{\mathit{eff}} = c_1 P \left(\overline{\theta^2}\right)^3 \qquad (26)$$

The numerical coefficient $c_1 = 0.05082$ differs by less than 10% from the one in eq (13). This lends credence to the method adopted of reducing the Fokker-Planck eq (3) to the diffusion eq (13). In particular, the assumption of interpreting eq (6) as being locally valid along the whole chain seems warranted. Thus, this justifies my use of an angular distribution independent of end effects.

The longitudinal dispersion of long DNA boils down to a computation of the second moment $\overline{\theta^2}$. Previously, we proposed a procedure which maps the statistics of the wormlike chain in a hard-walled nanochannel onto that of a chain confined in a harmonic potential [15]. This uses a coefficient in the free energy derived numerically for the polymer in the former case [20]. This yields [15]

$$\overline{\theta^2} = c_2 \left(\frac{A}{P}\right)^{2/3} \qquad (27)$$

where $c_2$ is estimated to be equal to 0.34. A fully numerical analysis of $c_2$ has also been presented recently: $c_2 = 0.3655$ [18]. These two numbers are so close that the harmonic approximation would seem to be entirely corroborated. Yang et al. [18] also computed $c_3 = 0.3402$ in the case where the nanochannel is purely cylindrical with $A$ the diameter.

Next, I compare the longitudinal dispersion derived here with experiment. In practice, the nanochannels used are not square but often rectangular (sides $A_1 \times A_2$). An extension of eq (9),(26) and (27) to the rectangular case proves that

$$\begin{aligned} D_{\mathit{eff}} &= \tfrac{1}{4} c_1 c_2^{\,3} (A_1 + A_2)^2 + P^{-1} \\ &= 0.000620 (A_1 + A_2)^2\, P^{-1} \end{aligned} \qquad (28)$$

to a very good approximation. Köster et al. studied the thermal fluctuations of F-actin filaments in microchannels by fluorescence microscopy [21,22]. In particular, they determined the distribution $\overline{G}(z)$ quantitatively which they compared with a numerical evaluation of a summation formula derived by Levi and Mecke [23]. In the latter theory, a variable analogous to $b$ occurs which is used as a fitting parameter but, unfortunately, the relation between $b$ and the size of the microchannel is not discussed. The difficulty is that analytical insight into the behavior of the distribution is hard to achieve via the Levi-Mecke approach (although their expression does go beyond the Gaussian approximation deduced here). Three of the curves for $\overline{G}(z)$ in Fig.3 of ref.[21] are close to Gaussian. From eq (15) the chain extension at midheight should be given by

$$\Delta z^2 = 2.76\, D_{\mathit{eff}} L \qquad (29)$$

The F-actin has a contour length L = 21μm and a persistence length P = 13μm. Values of $\Delta z$ and $D_{\mathit{eff}}$ are presented in as a function of the microchannel dimensions. Clearly, $D_{eff}$ is significantly underestimated by the present theory. On the other hand, in view of the sixth power law (eq (26)), the typical angle of the F-actin with respect to the center axis of the microchannel is underestimated by merely a factor of 1.2-1.3. Accordingly, it is entirely conceivable that the discrepancy is due to the limitations of the harmonic approximation; the real orientational distribution could be broader than Gaussian. The longitudinal dispersion of λDNA in a rectangular channel was studied earlier [16] but the nanochannel is relatively broader $\left(\langle\theta^2\rangle = O(1)\right)$ so a comparison with theory is really tentative. The quadratic dispersion $(\Delta z^2)$ is indeed found to be proportional to $L$ with a coefficient $D_{eff}$ of about 18nm. This is quite larger than the theoretical prediction of 0.77nm based on P=58nm for DNA in a 100x200nm$^2$ nanochannel. The typical orientation of the DNA is underestimated by a factor of 1.7. Tegenfeldt et al. [16] argue that a Daoud-de Gennes blob model may hold even under there conditions ($A_1,A_2=O(P)$). However, other intermediate regimes have been proposed and the subject is still under investigation [24].

Recent complete simulations [25] do seem to bear out that the present theory underestimates the longitudinal dispersion of stiff chains in cylindrical channels. At small diameters the distribution $\overline{G}(z)$ is indeed close to a Gaussian but the typical angle needs to be about 1.28 times larger in order to explain the dispersion (using eq (27) with coefficient $c_3$).

On the whole, the longitudinal dispersion of nanoconfined chains within the harmonic approximation underestimates the actual dispersion to a significant degree. But in terms of the typical orientation of the chain, the predictive value is off by a factor of about 1.3. Simulations of the orientational distribution across a channel show that it is close to but not precisely a Gaussian [18]. It is possible that an investigation of simulated distributions may explain this factor.

The extension $\Delta z$ at midheight is calculated from the curves in Fig. 3 of ref. [21]. The experimental $D_{eff}$ is given by eq (29). The theoretical $D_{eff}$ is computed via eq (28).

**Table I. Comparison of theory with experiment for F-actin**

| $A_1 \times A_2$ (μm$^2$) | $\Delta z / L$ | $D_{eff}$ (exp) (nm) | $D_{eff}$ (theory) (nm) |
|---|---|---|---|
| 1.4x1.5 | 0.015 | 1.7 | 0.32 |
| 1.4x4.0 | 0.02 | 3.0 | 1.12 |
| 1.4x5.8 | 0.03 | 6.8 | 2.0 |

[1] N. Douville, D. Huh, and S. Takayama, Anal. Bioanalyt. Chem. **391**, 2395 (2008).
[2] C. C. Hsieh and P. S. Doyle, Korea-Australia Rheol. J. **20**, 127 (2008).
[3] J. H. Kim, V. R. Dukkipati, S. W. Pang, and R. G. Larson, Nanoscale Res. Lett. **2**, 185 (2007).
[4] R. Riehn, W. Reisner, J. O. Tegenfeldt, Y. M. Wang, C. K. Tung, S. F. Lim, E. Cox, J. C. Sturm, K. Morton, S. Y. Chou, and R. H. Austin, in *Integrated Biochips for DNA Analysis*, R. H. Lim and A. P. Lee ed. (Springer, New York, 2007).
[5] G. B. Salieb-Beugelaar, K. D. Dorfman, A. van den Berg, and J. C. T. Eijkel, Lab Chip **9**, 2508 (2009).
[6] D. Stein, M. van der Heuvel, and C. Dekker, in *Nanofluidics: Nanoscience and Nanotechonology*, B. Edel and A. J. de Mello ed. (Royal Society of Chemistry, Cambridge, U.K., 2009).
[7] H. Yamakawa, *Helical Wormlike Chains in Polymer Solutions*. (Springer, Berlin, 1997).
[8] H. A. Kramers, Physica **7**, 284 (1940).
[9] V. G. Levich, *Physicochemical Hydrodynamics*. (Prentice-Hall, Englewood Cliffs, N.J., 1962).
[10] G. Taylor, Proc. Roy. Soc. A **219**, 186 (1953).
[11] G. Taylor, Proc. Roy. Soc. A **225**, 473 (1954).
[12] K. M. Case, Ann. Phys. **9**, 1 (1960).
[13] J. H. Tait, *Introduction to Neutron Transport Theory*. (Longmans, 1964).
[14] T. Odijk, Macromolecules **16**, 1340 (1983).
[15] K. Jo, D. M. Dhingra, T. Odijk, J. J. de Pablo, M. D. Graham, R. Runnheim, D. Forrest, and D. C. Schwartz, Proc. Acad. Sci. USA **104**, 2673 (2007).
[16] J. O. Tegenfeldt, C. Prinz, H. Cao, S. Chou, W. W. Reisner, R. Riehn, Y. M. Wang, E. C. Cox, J. C. Sturm, P. Silberzan, and R. H. Austin, Proc. Acad. Sci. USA **101**, 10979 (2004).
[17] T. W. Burkhardt, J. Phys. A. **28**, L629 (1995).
[18] Y. Yang, T. W. Burkhardt, and G. Gompper, Phys. Rev. **76**, 011804 (2007).
[19] T. Odijk, J. Chem. Phys. **125**, 204904 (2006).
[20] T. W. Burkhardt, J. Phys. A. **30**, L167 (1997).
[21] S. Koster and T. Pfohl, Cell Motility Cytoskeleton **66**, 771 (2009).
[22] S. Koster, D. Steinhauser, and T. Pfohl, J. Phys. Cond. Mat. **17**, S4091 (2005).
[23] P. Levi and K. Mecke, Europhys. Lett. **78**, 38001 (2007).
[24] T. Odijk, Phys. Rev. E **77**, 060901 (R) (2008).
[25] P. Cifra, Z. Benkova, and T. Bleha, J. Phys. Chem. B **112**, 1367 (2008).